\documentclass[a4paper,10pt]{article}
\usepackage[utf8]{inputenc}

\title{ Deformation of the  Dirac Equation   }
\author{Mir Faizal$^1$ and  Sergey I. Kruglov$^2$  \\ $^1$Department of Physics and Astronomy, \\  University of Waterloo,   Waterloo,\\
Ontario N2L 3G1, Canada  
\\
$^2$Department of Chemical and Physical Sciences,\\ University of Toronto,
3359 Mississauga Rd. North, \\
Mississauga, Ontario,  L5L 1C6, Canada
}
\date{}
\begin{document}

\maketitle

\begin{abstract}
In this paper, we will first clarify the physical meaning of having a minimum measurable time. 
Then we will combine  the deformation of the  Dirac equation   due to the existence of minimum measurable 
length and time scales   with its deformation 
due  to the doubly special relativity.  
  We will also analyse this deformed  Dirac equation in curved spacetime, and observe that 
  this deformation of the Dirac equation also leads to a  non-trivial modification  of   general relativity. 
Finally, we  will  analyse the stochastic quantization of this deformed Dirac equation on curved spacetime. 
\end{abstract}
%\section{}

\section{Introduction}

All most all approaches to quantum gravity predict the  existence of a minimum measurable length scale. 
For example, in string theory the smallest probe that can be used for analysing spacetime is the fundamental string.
So, it is not possible to probe spacetime at length scales smaller than the string length scale. Thus, string theory
comes naturally equipped with a minimum length scale, which is the string length scale  \cite{z2}-\cite{2z}.
A minimum length scale also exists in loop quantum gravity. In fact, this minimum length scale in loop quantum gravity turns 
the big bang into a big bounce \cite{z1}.
The existence of a minimum length scales can also be inferred  from black hole physics \cite{z4}-\cite{z5}.
This is because the energy required to probe spacetime at length scales smaller than Planck length is more than the energy required 
to form a black hole in that region of spacetime. Thus, it can be inferred from black hole physics that any 
theory of quantum gravity should naturally come equipped with an  minimum measurable length scale of the order of Planck scale.

The existence of a minimum length is not consistent with the usual  Heisenberg uncertainty principle.
This is because according to the usual  Heisenberg uncertainty principle, it is possible to  measure
length with arbitrary accuracy if momentum is not  measured. So, the usual  Heisenberg uncertainty principle
 has to be generalized to generalized uncertainty principle, to make the existence of  minimum length
consistent with quantum mechanics \cite{1}-\cite{15}. However, the Heisenberg uncertainty principle is closely related to
the Heisenberg algebra, and so,  the deformation of the usual Heisenberg uncertainty principle, will also lead to a deformation of the 
Heisenberg algebra \cite{17}-\cite{53}. This  deformation of the Heisenberg algebra changes the  momentum operator in the coordinate
representation, and this will in turn produce correction terms for all quantum mechanical systems \cite{18}-\cite{54}.

A different form of  deformation of the Heisenberg algebra occurs due a modification of the special relativity to doubly special relativity 
 \cite{2}-\cite{3}. Apart from the velocity of light, the Planck energy is also a universal constant in  doubly special relativity .
The doubly special relativity  is motivated by 
a modification of the usual energy-momentum dispersion relation, which in turn is motivated by studies done on discrete spacetime \cite{1q}, 
the spontaneous symmetry breaking of
Lorentz invariance in string field theory \cite{2q}, spacetime foam models \cite{3q}, spin-network in loop quantum gravity \cite{4q}, 
non-commutative geometry \cite{5q}, and Horava-Lifshitz gravity \cite{6q}.
In fact, it  has also been possible to generalize doubly special relativity to doubly general relativity,
and the resultant theory is know as gravity's rainbow \cite{n1}-\cite{n2}. 
The deformations of the Heisenberg algebra caused by the existence of a minimum measurable length
and   doubly special relativity have been combined into  a single deformation of
the Heisenberg algebra \cite{n4}-\cite{n5}. Corrections to various  physical systems have been studied using this  deformed Heisenberg algebra. 
For example, the  transition rate of ultra cold neutrons in gravitational field
received corrections due to this deformed Heisenberg algebra \cite{n6}. Similarly, the correction terms for the
Lamb shift and the Landau levels have also been calculated \cite{n7}. One of the most interesting consequence of this deformed algebra is 
that it leads to a discretization of space \cite{n4}.
This deformation of the Heisenberg algebra changes the form of the first quantized field theories, 
and second quantization of these  deformed field theory has also been studied  \cite{n9}-\cite{mir}.
The deformation of the Wheeler-DeWitt equation corresponding to this algebra has   
been performed   \cite{mir2}. It was observed that due to such a deformation the big bang singularity can be avoided. This happens because there is 
a minimum value for the scale factor of  the universe. So, 
 the  universe can not become smaller than this minimum value, and hence,   the big bang singularity is avoids.  

It may be noted that it is possible to define a system with minimum measurable time, and if the principle of covariance in relativity is taken 
seriously, then a minimum measurable length scale should also correspond to a minimum measurable time scale.  
This can be  done by defining the observable time  
as the time of occurrence of an certain type of event associated with the 
 appearance of some specified value of a dynamical quantity \cite{time}-\cite{time1}.
 In fact, there are good reasons to treat the space and time on equal footing, as it has been argued  that 
 by treating the space and time on equal footing  the  black-hole information paradox gets naturally resolved \cite{blip}-\cite{blip1}.
 Thus, it is possible to generalize the work on deformation of Heisenberg algebra based on 
 minimum length scale  to  both  minimum length and  time scales.  
 It may also be noted that another interesting deformation of quantum mechanics comes from 
 stochastic quantization  \cite{stochastic}-\cite{stochastic1}. 
 In fact,  it has been argued that the Gell-Mann–Oakes–Renner   relation has
  an analogous meaning in QCD, and this leads to  the existence of a Thouless energy in QCD \cite{qcd}-\cite{qcd1}.
 It has been possible to calculate this Thouless energy in QCD in the framework of 
  stochastic quantization \cite{qcd2}. 
  General relativity can be treated as gauge theory \cite{gauge11}-\cite{gauge44}, and so, it is possible to apply the results obtained 
  for   stochastic quantization of gauge theories to gravity. 

In this paper, we will first clarify what it means to have a minimum measurable time in a theory.
This will be done by defining an minimum measurable 
value to the certainty with which the  occurrence of an event of a certain type can be measured by 
  the appearance of some specified value of a dynamical quantity. 
Then we will combine the  deformation of the Heisenberg algebra caused by the   doubly special relativity, with its deformation 
caused by  
existence of minimal measurable length and time scales. 
We will also complete this algebra by including a covariant deformation of its temporal part. 
The Dirac equation consistent with this  deformed algebra Heisenberg will contain 
non-local fractional derivative terms. 
However, such terms will be given a formal meaning in the framework of harmonic extension of functions.
We will also analyse the effect this deformation of Dirac equation has on general relativity.
We observe that apart from deforming the matter action, this deformation also 
leads to a non-trivial modification of the gravitational part of the action.
As general relativity can be viewed as a gauge theory, we will analyse the deformed Dirac equation on curved spacetime
in the framework of stochastic quantization. 
We will perform our analysis semi-classically taking  the gravitational field 
  as a fixed background. 
We will also demonstrate that this stochastic quantization of the deformed Dirac equation can be done in the superspace formalism.

\section{Deformation of Heisenberg Algebra}

In this section, we first generalize the deformation of quantum mechanics caused by the existence of a minimum measurable length scale to 
the deformation caused by the existence of both  minimum measurable length and time scales.  
The existence of minimum length causes the following deformation of the uncertainty principle \cite{17}-\cite{53}, 
\begin{equation}
\Delta x \Delta p = \frac{\hbar}{2} [1 + \beta (\Delta p)^2],
\end{equation}
and this in turn deforms the Heisenberg algebra to
\begin{equation}
[x^i, p_j ] =
i \hbar [\delta_{j}^i + \beta p^2 \delta_{j}^i + 2 \beta p^i p_j],
\end{equation}
where  $\beta = \beta_0 \ell_{Pl}^2/ \hbar^2 $,   $\beta_0$ is a constant normally assumed to be of order one,
and  $\ell_{Pl} \approx 10^{-35}~m$.  The coordinate representation of the momentum operator can now be written as
\begin{equation}
 p_i = i \hbar \partial_i  (1 -  \beta  \hbar^2 \partial^i \partial_i  ).
\end{equation}
Now a deformation of the Heisenberg algebra corresponds to the existence of a 
minimum measurable time is also expected to take place due to principles of covariance in relativity. Thus, we expect the
uncertainty between energy and time to get deformed to
\begin{equation}
\Delta E \Delta t = \frac{\hbar}{2} [1 +\beta  (\Delta E)^2],
\end{equation}
where we have set $ t \to i t$. 
It may be noted that such a deformation has already been studied \cite{r}.  
This deformation will in turn changes the representation of the Hamiltonian operator to
\begin{equation}
 H = i \hbar\partial_t (1 - \beta \hbar^2   \partial_t^2 ).
\end{equation}
It may be noted that it is possible to define time as an observable in quantum mechanics, 
with reference to the evolution of some 
non-stationary quantity \cite{time}-\cite{time1}. 
Thus, the observable time  
is the time of occurrence of an event of a certain type,
defined by the appearance of some specified value of a dynamical quantity.
So, by viewing time as an observable in quantum mechanics, 
it is possible to give physical meaning to the commutator $[H, t] = i\hbar$. 
Now the deformation of the Hamiltonian operator caused by the existence of a minimum measurable time scale corresponds to 
the deformation of the commutator to $[H, t] = i\hbar [1 + 3 \beta  H^2 ] $, where again we have set $ t \to it$.
This would physically  correspond to the case 
where there is   an minimum measurable 
value to the certainty with which the  occurrence of an event of a certain type can be measured by 
  the appearance of some specified value of a dynamical quantity. 

Now we will 
analyse the deformation of the  Heisenberg algebra which incorporates both the deformations corresponding 
to the existence of minimum length and   doubly special relativity. 
The deformation of  the Heisenberg algebra caused  doubly special relativity is given by 
\begin{equation}
 [x^i, p_j] =
i  \hbar[1 - \tilde \beta |p^k p_k|^{1/2} \delta_{j}^i + \tilde \beta^2  p^i p_j],
\end{equation}
where  $\tilde \beta = \ell_{Pl} \approx 10^{-35}~m$.
It is possible to combine the deformation occurring due to the existence of a minimum length scale and
the deformation occurring due to doubly special relativity into a single deformation of the Heisenberg algebra
\cite{n4}-\cite{n5}
\begin{eqnarray}
 [x^i, p_j] &=&  i \hbar \left[  \delta_{j}^i - \alpha |p^k p_k|^{1/2} \delta_{j}^i + \alpha |p^k p_k|^{-1/2} p^i p_j
 \right.  \nonumber \\  && \left. + \alpha^2 p^k p_k \delta_{j}^i + 3 \alpha^2 p^i p_j\right],
\end{eqnarray}
where $M_{Pl}$ is the Planck mass, $\alpha = {\alpha_0}/{M_{Pl}c} = {\alpha_0 \ell_{Pl}}/{\hbar}$,
and $M_{Pl} c^2 \approx 10^{19}~GeV$ is the Planck energy. 
It has been suggested that the parameter $\alpha_0$   could
be situated at an    intermediate length
scale between Planck scale and electroweak scale \cite{n7}. 
If that is the case, then it will be possible to device experiments to analyse the phenomenological effects 
of quantum gravitational phenomena. 
In fact, in that case,  it might be possible to observe the 
upper bound for $\alpha_0$ by using   quantum optics techniques \cite{optics}.  
The coordinate representation of the momentum operator can now be written as
\begin{equation}
p_i = -i \hbar\left(1 -  \hbar\alpha \sqrt{- \partial  ^j \partial_j} - 2\hbar^2 \alpha^2    \partial ^j
\partial _j\right) \partial _i.
\label{3}
\end{equation}
Motivated by the deformation of the momentum operator  corresponding to the existence of a minimal measurable time,
 we assume  the following form for the 
 deformed  four dimensional momentum operator in the coordinate representation, 
\begin{equation}
p_\mu = -i \hbar \left(1 -  \hbar \alpha \sqrt{- \partial  ^\nu \partial_\nu} - 2 \hbar^2 \alpha^2  \partial ^\nu
\partial _\nu\right) \partial _\mu.
\end{equation}
This deformed momentum operator statisfies the following algebra, 
\begin{eqnarray}
 [x^\mu, p_\nu] &=& i \hbar \left[  \delta_{\nu}^\mu - \alpha |p^\rho p_\rho|^{1/2} \delta_{\nu}^\nu + \alpha |p^\rho p_\rho|^{-1/2} p^\mu p_\nu
\right.  \nonumber \\  && \left. + \alpha^2 p^\rho p_\rho \delta_{\nu}^\mu + 3 \alpha^2 p^\mu p_\nu\right].
\end{eqnarray}
where again  we have set $ t  \to i t$. 
Here again  time has to be viewed as an  observable, for the  temporal part of this commutator to be 
physically meaningful.  
This is again  done by first defining  the observable time  
as the time of occurrence of an event of a certain type at which   
  some specified value gets assigned to a dynamical quantity  \cite{time}-\cite{time1}.
Then viewing this algebra as  the existence of a  minimum measurable 
value to the certainty with which the  occurrence of that  event   can be measured by 
  the appearance of this specified value for that  dynamical quantity. 
It may be noted that if time is taken as an observable, then
motivated by the principle of covariance of space and time, it was expected that this deformed algebra should also  included 
its temporal part.  
We will assume   this to be a general form for the momentum operator in all field theories. Thus, we will be able to 
analyse the deformation of the Dirac equation by using this  deformed  momentum operator. 
We will also  chose natural unites and set $ c = \hbar = 1 $.
\section{Deformed Dirac Equation }

In the previous section, we have analysed the deformation of the  Heisenberg algebra caused the doubly special relativity 
along with the existence of minimum measurable length and time scales.
In this section, we will analyse the Dirac equation corresponding to this deformation of the Heisenberg algebra.
Now we can write the Dirac equation corresponding to this deformed Heisenberg algebra as follows \cite{n9}-\cite{mir}, 
\begin{equation}
i \gamma^\mu\left(1-\alpha\sqrt{ - \partial^\nu   \partial _\nu}- 2\alpha^2  \partial ^\nu   \partial _\nu\right)  \partial _{\mu}\psi(x)+m\psi(x)=0.
\end{equation}
where   $\gamma^\mu$ are Dirac's matrices obeying $\gamma^\mu\gamma^\nu+\gamma^\nu\gamma^\mu=2\eta^{\mu\nu}$, and 
$\psi(x)$ is bispinor \cite{Ahieser}.

This Dirac equation  contains non-local fractional derivative term, however, this term can be given a formal
meaning in the framework of harmonic extensions of a function   \cite{lifz2}-\cite{lifz}.
We start by defining a harmonic extension of  $ \psi   : R^n \to R$, which is a
harmonic function $u: R^n \times (0, \infty) \to R$. The    restriction  of $u$ to  $R^n$  coincides with
 $ \psi   : R^n \to R$.
We can also define $\partial^2_{n+1}$ as the Laplacian in $R^{n+1}$, such that  $x \in R^n$ and $ y \in R$.
Now for a given  $\psi$, we can solve the Dirichlet problem given by $u(x, 0) = \psi (x)$ and $\partial^2_{n+1} u (x, y)=0$, to find $u$. 
Thus, for a given smooth function $C^\infty_0 (R^n) $, there is a unique harmonic
extension  $ u \in C^\infty (R^n \times (0, \infty))$.

A formal meaning can now be given to $\sqrt{-\partial^\mu\partial_\mu}$ by analysing its action on  $\psi: R^n \to R $. 
We require in that case the harmonic extension of the $\psi$ to satisfy $u: R^n \times (0, \infty) \to R$,
\begin{equation}
\sqrt{-\partial^\mu\partial_\mu}\psi (x) = - \frac{\partial u (x, y)}{ \partial y}\left. \right| _{y =0}.
\end{equation}
Here $u_y(x, y)$ is the harmonic extension of $\sqrt{-\partial^\mu\partial_\mu}\psi (x)$ to $R^n \times (0, \infty)$, as $u: R^n \times (0, \infty) \to R$ is a harmonic extension of   $ \psi   : R^n \to R $. The repeated  application  of $\sqrt{-\partial^\mu\partial_\mu}$ on $\psi$ can be written as
\begin{eqnarray}
\left[\sqrt{-\partial^\mu\partial_\mu}\right]^2\psi (x) &=& \frac{\partial^2 u(x, y)}{\partial y^2}\left. \right|_{y =0}
\nonumber \\
&=& - \partial ^2_n u(x, y)\left. \right|_{y =0}.
\end{eqnarray}
This equation gives a formal meaning to $ \sqrt{- \partial^\mu \partial_\mu }$.
The harmonic extension of $\partial_\mu
\psi (x) $  can be written as
$ \partial_\mu u (x, y)$ because  $u (x, y)$ is the harmonic extension
of the  $\psi (x)$, and  $\sqrt{-\partial^\mu\partial_\mu} \psi = - u_y (x, 0) $. Now as  $u \in C^2 (R \times (0, \infty))$, so, we can write
\begin{equation}
\sqrt{-\partial^\mu\partial_\mu} \partial_\nu    \psi (x)  = - \partial_\nu u_y (x, y) \left. \right|_{y =0}.
\end{equation}
Thus, any derivative commutes with this non-local operator $\sqrt{-\partial^\mu\partial_\mu}$,
\begin{equation}
\sqrt{-\partial^\mu\partial_\mu} \partial_\nu    \psi (x)  = \partial_\nu \sqrt{-\partial^\mu\partial_\mu}    \psi (x) .
\end{equation}

Now we have given a formal meaning to the  non-local Dirac equation with fractional derivative terms. We can also write this 
deformed Dirac equation in a matrix valued
formalism. To do that we introduce the following functions,
\begin{eqnarray}
 \chi(x)&=& - \frac{1}{m}\partial_5\psi(x),  \\ \nonumber
 \phi(x)&=&-\frac{1}{m^2}\partial^\mu\partial_\mu \psi(x)
\\ \nonumber &=& -\frac{1}{m}\partial_5\chi,
\end{eqnarray}
 where
$\partial_5\equiv \sqrt{- \partial^\mu \partial_\mu }$. Then this deformed Dirac equation  can be written as
\begin{eqnarray}
 i \gamma^\mu\left(\partial_{\mu}\psi(x)+\alpha m\partial_\mu\chi(x)+ 2\alpha^2 m^2\partial_\mu\phi(x)\right)+m\psi(x)&=&0, \nonumber \\
\partial_5\psi(x)+m\chi(x)&=&0,\nonumber \\
 \partial_5\chi(x)+m\phi(x)&=& 0.
\end{eqnarray}
We introduce a 12-component wave function, $\Psi (x) = (\psi (x), \chi (x), \phi (x) )$, and
write the deformed  Dirac equation as
\begin{equation}
i \Gamma^\mu\partial_\mu\Psi(x)+i \Gamma^5\partial_5\Psi(x)+m\Psi(x)=0,
\label{8}
\end{equation}
where
\begin{eqnarray}
  \Gamma^\mu=\gamma^\mu\otimes A, &&
  \Gamma^5=I_4\otimes B, \nonumber \\
  A=\varepsilon^{1,1}+\alpha m\varepsilon^{1,2}+2\alpha^2 m^2\varepsilon^{1,3},  &&
  B=\varepsilon^{2,1}+\varepsilon^{3,2}.
\end{eqnarray}
Here   $A$ is a projection matrix $A^2=A$,  $B^3=0$, and
$I_4$ is a unit $4\times 4$-matrix. Thus, advantage of using this formalism is that the non-local fractional derivative term
can be effectively written as a local term. Thus, using 
the notations $\Gamma^A=(\Gamma^\mu,\Gamma^5)$, and 
$\partial_A=(\partial_\mu,\partial_5)$, we can now write the deformed  Dirac equation as
\begin{equation}
\left(i \Gamma^A\partial_A+m\right)\Psi(x)=0,
\label{10}
\end{equation}
where $A=1,2,3,4,5$. It may be noted that this is effectively written as a local Dirac equation. 
Now in the momentum space we have the following result
\begin{equation}
\left( \mathcal{P} - m\right)\Psi(p)=0,
\label{11}
\end{equation}
where $\mathcal{P} =\Gamma^\mu p_\mu +\Gamma^5 p_5$,   and so,
\begin{equation}
\mathcal{P} =\left(\begin{array}{ccc}
            \gamma^\mu p_\mu & \alpha m \gamma^\mu p_\mu & 2\alpha^2 m^2\gamma^\mu p_\mu \\
            p_5I_4 & 0 & 0 \\
            0 & p_5I_4 & 0
          \end{array}
          \right).
\label{12}
\end{equation}
We also have $A^2=A, B^2=\varepsilon^{3.1}, B^3=0$,  and
 \begin{eqnarray}
  \Gamma^\mu\Gamma^\nu+\Gamma^\nu\Gamma^\mu &=& 2\eta^{\mu\nu}I_4\otimes A, \nonumber \\
  \Gamma^\rho\left(\Gamma^\mu\Gamma^\nu+\Gamma^\nu\Gamma^\mu\right)&=& 2\eta^{\mu\nu}\Gamma^\rho.
 \end{eqnarray}

\section{Deformation  of General Relativity}
It is possible to write the deformation of any gauge theory by converting the deformed derivatives into 
deformed covariant derivatives \cite{skdj}-\cite{n}. In fact,  for a Lie algebra $[T_A, T_B] = i f_{AB}^C T_C$, the gauge 
field $ A_\mu = A^A_\mu T_A$  can be used to write a covariant derivative $D_\mu = \partial_\mu + i A_\mu$.
It has been demonstrated that 
it is possible to analyse a deformation of a gauge theory by deforming 
$ -i   \left(1 -  \alpha \sqrt{- \partial  ^\nu \partial_\nu} - 2   \alpha^2  \partial ^\nu
\partial _\nu\right) \partial _\mu $ to $ -i   \left(1 -  \alpha 
\sqrt{- D  ^\nu D_\nu} - 2   \alpha^2  D^\nu
D _\nu\right) D_\mu $ \cite{mir}. In this paper, we analyse such a deformation of general relativity. 
It is possible formulate  the general relativity as a gauge theory by considering the translation group as a gauge group 
\cite{gauge11}-\cite{gauge22}.  
Even though the torsion plays a central role in this formalism,  it can be demonstrated that this formalism 
is equivalent to general relativity. 
It is also possible to  formulate  the general relativity as a gauge theory by using the Lorentz group as a gauge group 
\cite{gauge44}. We expect that the gauge theories to get modified  due to the deformation of the Heisenberg algebra, 
so, the deformation of the  Heisenberg algebra will also lead to a non-trivial modification of general relativity. 
In this paper, we will analyse general relativity as a gauge theory whose gauge group is the  Lorentz group. 
Now under the Lorentz transformation   given by 
$ x^a \to x^b \Lambda^a_b
$,    a fermion will transform  as 
\begin{equation}
 \psi  \to U \psi = \exp \left(\frac{i}{2}\Lambda^{ab}(x) \Sigma_{ab} \right),
\end{equation}
where $\Sigma_{ab}$ describes the generators of Lorentz group represented within Dirac spinor space as 
$\Sigma_{ab} = -i [\gamma_a, \gamma_b]/4$. 
Now we define 
$
 g_{\mu\nu} e^\mu_a e^\nu_b = \eta_{ab}, 
$ and  so, we can write a spinor covariant derivative as 
\begin{equation}
 D_a = e^\mu_a (\partial_\mu I + \frac{i}{2} \omega^{ab}_\mu \Sigma_{ab})= e^\mu_a D_\mu. 
\end{equation}
Here $\omega_\mu^{ab}$ is the spin connection which is given by 
\begin{eqnarray}
 \omega_\mu^{ab} &=& 2 e^{\nu a}\partial_\mu e_\nu^b - 2 e^{\nu b} \partial_\mu e_\nu^a - 2 e^{\nu a}
 \partial_\nu e_\nu^b + 2 e^{\nu b}\partial_\nu e_\mu^a\nonumber \\ && + e_{\mu c}e^{\nu a} e^{\rho b}\partial_\rho e^a_\nu 
 - e_{\mu c} e^{\nu a} e^{\rho b} \partial_\nu e^c_\rho. 
\end{eqnarray}
The commutator of these two covariant derivative can be written as 
\begin{equation}
 [D_\mu, D_\nu] = A^c_{\mu \nu}D_c + \frac{i}{2}R^{ab}_{\mu\nu} \Sigma_{ab}. 
\end{equation}
We can set the torsion  to zero, $A^c_{\mu\nu} =0$, and write the Riemann tensor with two 
Minkowshi indices   as 
\begin{equation}
 R^{ab}_{\mu\nu} = \partial_\mu \omega_\nu^{ab} - \partial_\nu \omega_\mu^{ab} + 
 \omega^{ac}_\mu \omega^{b}_{c\nu} -  \omega^{ac}_\nu \omega^{b}_{c\mu}.
\end{equation}
Here the spin connection  $\omega_\mu^{ab}$ transforms as 
\begin{equation}
 \omega_\mu^{ab} \to \left[ U  \omega_\mu U^{-1} -  (\partial_\mu U)  U^{-1}  \right]^{ab}, 
\end{equation}
so, $ R^{ab}_{\mu\nu}$ transforms as 
\begin{equation}
  R^{ab}_{\mu\nu} \to [U]^a_c R_{\mu\nu}^{cd} [U^{-1}]^{b  }_{d }. 
\end{equation}
Now we can define the scalar curvature  as $ R =   e_a^\mu e_b^\mu R^{ab}_{\mu\nu} $, and 
  write the  action, with $\kappa = 8\pi G $, as  
\begin{equation}
 S = \frac{1}{2 \kappa} \int d^4 x e R  + \int d^4 x e \bar \psi  (i \gamma^a e_a^\mu D_\mu -m ) \psi.  
\end{equation}
So, the Einsteins equation can now be written as 
\begin{equation}
 R_{\mu}^a - \frac{1}{2} e^a_\mu R = \kappa T^a_\mu, 
\end{equation}
where $T^a_\mu$ is the energy momentum tensor corresponding to the Dirac equations. 

Now we can write the  deformed  Dirac equation in curved spacetime by using the assumption    that the 
form of the coordinate  representation of the deformed momentum operator will be the same in all 
field theories. So, we can replace the derivatives in the coordinate  representation of the deformed momentum operator 
by covariant derivatives. 
Thus,   the deformed  Dirac equation can be written as  \cite{n9}-\cite{mir}
\begin{equation}
\gamma^\mu\left(1-\alpha\sqrt{- D ^\nu  D _\nu} - 2\alpha^2 D ^\nu  D _\nu\right) D _{\mu}\psi(x)+m\psi(x)=0. 
\end{equation}
 Now if we define $\mathcal{D}_\mu =  (1-\alpha\sqrt{- D ^\tau   D _\tau } - 2\alpha^2 D ^\tau   D _\tau )D_\mu$, 
 then we can write the deformed commutator  as
\begin{eqnarray}
 [\mathcal{D}_\mu, \mathcal{D}_\nu] &=&  \mathcal{A}^c_{\mu \nu}D_c + \frac{i}{2}\mathcal{R}^{ab}_{\mu\nu} \Sigma_{ab} 
 \nonumber \\  &=& 
 A^c_{\mu \nu}D_c + \frac{i}{2}R^{ab}_{\mu\nu} \Sigma_{ab} 
  +  \alpha \left[   A(\alpha) ^{c  }_{\mu \nu}D_c + \frac{i}{2}{    R}(\alpha) ^{ab}_{ \mu\nu} \Sigma_{ab}   \right] \nonumber \\ && 
 + \alpha^2 \left[   {  A}(\alpha^2) ^{c  }_{\mu \nu}D_c + \frac{i}{2}  {   R}(\alpha^2) ^{ab}_{\mu\nu} \Sigma_{ab} \right]. 
\end{eqnarray}
So, we can write 
\begin{eqnarray}
 \mathcal{A}^c_{\mu \nu} =  {A}^c_{\mu \nu} +  \alpha {A}(\alpha )^c_{\mu \nu} + \alpha^2  {A}(\alpha^2 )^c_{\mu \nu}, 
 \nonumber \\ 
 \mathcal{R}^{ab}_{\mu\nu} = {R}^{ab}_{\mu\nu}  + \alpha {R}(\alpha )^{ab}_{\mu\nu} + \alpha^2 {R}(\alpha^2 )^{ab}_{\mu\nu},  
\end{eqnarray}
where  $  A(\alpha )  ^{c  }_{\mu \nu}D_c + {i}  {  R}(\alpha ) ^{ab}_{\mu\nu} \Sigma_{ab}/2 $ and
$   {  A}(\alpha^2) ^{c  }_{ \mu \nu}D_c +  {i}   {   R}(\alpha^2)^{ab}_{ \mu\nu} \Sigma_{ab}/2 $
are correction terms generated from the deformation of Heisenberg  algebra, 
 \begin{eqnarray}
  {A} ^c_{\mu \nu}  D_c + {R} ^{ab}_{\mu\nu}  \Sigma_{ab} &=&  \left[ D_\mu,  D_\nu \right],   
 \nonumber \\ 
 {A}(\alpha )^c_{\mu \nu}  D_c + {R}(\alpha  )^{ab}_{\mu\nu}  \Sigma_{ab} &=&  - \left[ \sqrt{- D ^\tau   D _\tau } D_\mu,   D_\nu \right]   
  - \left[ D_\mu,  \sqrt{ - D ^\tau  D _\tau }D_\nu \right],  
\nonumber \\ 
{A}(\alpha^2 )^c_{\mu \nu}  D_c + {R}(\alpha^2  )^{ab}_{\mu\nu}  \Sigma_{ab}
&=& -  2 \left[D^\tau   D_\tau D_\mu,  D_\nu \right] -2 \left[ D_\mu,  D ^\tau   D _\tau  D_\nu \right] \nonumber \\ && 
  +  \left[\sqrt{- D ^\tau   D _\tau }  D_\mu, \sqrt{ - D ^\rho   D _\rho } D_\nu \right]. 
\end{eqnarray}
We again set the torsion   equal to zero, $A^c_{\mu\nu} =  A(\alpha)^c_{\mu\nu} =  A(\alpha^2)^c_{\mu\nu} =0$, and 
write the  deformed   Riemann tensor with two 
Minkowshi indices   as 
\begin{equation}
 \mathcal{R}^{ab}_{\mu\nu} = \partial_\mu \omega_\nu^{ab} - \partial_\nu \omega_\mu^{ab} + 
 \omega^{ac}_\mu \omega^{b}_{c\nu} -  \omega^{ac}_\nu \omega^{b}_{c\mu} +   {R}(\alpha)^{ab}_{\mu\nu}  +  {R}(\alpha^2)^{ab}_{\mu\nu} 
\end{equation}
Now as  $ \mathcal{D}_\mu = (1-\alpha\sqrt{- D ^\nu  D _\nu} - 2\alpha^2 D ^\nu  D _\nu )  D_\mu $
also transforms as $D_\mu $, the transformation   $ \mathcal{R}^{ab}_{\mu\nu}$ can be written as 
\begin{eqnarray}
  \mathcal{R}^{ab}_{\mu\nu} &\to& [U]^a_c   \mathcal{R}_{\mu\nu}^{cd} [U^{-1}]^{b  }_{d }
  \nonumber \\ 
  &=&  [U]^a_c \left( {R}_{\mu\nu}^{cd} +  {R}(\alpha) _{\mu\nu}^{cd} +   {R}(\alpha^2)_{\mu\nu}^{cd} \right)[U^{-1}]^{b  }_{d }. 
\end{eqnarray}
Now we can   define the deformed  scalar curvature as $   \mathcal{R} =   e_a^\mu e_b^\mu   \mathcal{R}^{ab}_{\mu\nu} $, and 
  write the deformed  action as  
\begin{equation}
 S = \frac{1}{2 \kappa} \int d^4 x e   \mathcal{R}  + \int d^4 x e \bar \psi  (i \gamma^a e_a^\mu D_\mu -m ) \psi.  
\end{equation}
The deformed  Einsteins equations can also be written as 
\begin{equation}
 \mathcal{R}^a_\mu - \frac{1}{2} e^a_\mu \mathcal{R} = \kappa \mathcal{T}^a_\mu, 
\end{equation}
where $\mathcal{R}^a_\mu = \mathcal{R}^{ab}_{\mu\nu} e^b_\nu$ is the Ricci tensor generated from the deformed 
Riemann tensor,  and $\mathcal{T}^a_\mu = T^a_\mu + \alpha T(\alpha)^a_\mu + \alpha^2 T(\alpha^2)^a_\mu $
is the deformed energy momentum tensor. 
It may be noted that apart from the deformation of the energy-momentum tensor corresponding to the 
deformation of the original theory, the gravitational action  for this theory also gets deformed. 

\section{Stochastic Quantization}
In this section, we will analyse the stochastic quantization of Dirac equation in curved spacetime. We 
will analyse this model semi-classically and treat gravity as a fixed classical background and 
analyse the stochastic quantization of Dirac equation on this fixed gravitational background. 
 Stochastic quantization is performed by introducing an extra fictitious time variable $\tau$, 
 and solving the corresponding Langevin equations. Thus, we can start by introducing an extra time 
 variable for the fermionic fields $\psi (x) \to \psi (x, t), \bar\psi (x) \to \bar\psi (x, t) $, 
 and then writing down the Langevin equations \cite{stochastic}-\cite{stochastic1}. 
As the action in these Langevin equations is a deformed Dirac action, 
we have to include an appropriate Kernel $K (x, y)$ to ensure the relaxation 
 process is such that the systems will approach equilibrium as $\tau \to 0$. 
We also  need to introduce $\eta (x, \tau)$ and $\eta  (x, \tau)$ as 
 anticommuting fermionic Gaussian noise. Here  $\eta (x, \tau)$ and $\eta  (x, \tau)$ satisfy 
 \begin{eqnarray}
   \langle \eta(x, \tau) \rangle &=& 0 
 \nonumber \\ \langle \bar \eta (x, \tau) \rangle &=& 0, \\ \nonumber
  \langle \eta  (x', \tau')  \bar \eta (x, \tau) \rangle &=& 2 \delta(\tau - \tau' ) K(x, x') \delta^4 (x - x'). 
 \end{eqnarray}
 Now using this  anticommuting fermionic Gaussian noise, we can write the Langevin equations as 
 \begin{eqnarray}
  \frac{\partial \psi (x, \tau) }{\partial \tau } &=& - \int d^4 y K(x, y)  \frac{\delta S  [\psi, \bar \psi ]}{ \delta \bar \psi } + \eta (x, \tau),
  \\ \nonumber
    \frac{\partial \bar \psi (x, \tau) }{\partial \tau } &=&  \int d^4 y \frac{\delta S  [\psi, \bar \psi ]}{ \delta   \psi }K(x, y)
    + \bar \eta (x, \tau), 
 \end{eqnarray}
 where the action for the deformed Dirac equation on curved spacetime is given by 
\begin{equation}
 S [\psi, \bar \psi]  = \int d^4 x e  \bar \psi [ 
 i \gamma^\mu\left(1-\alpha\sqrt{ - D^\nu   D _\nu}- 2\alpha^2  D ^\nu   D _\nu\right)  D _{\mu}- m] \psi. 
\end{equation}
As we are analysing the system semi-classically, our dynamical variables are only $\psi$ and $\bar \psi $.
The partition function for the deformed Dirac equation can be written as 
 \begin{equation}
  Z = \int D \bar \eta D \eta  \exp \left( - \frac{1}{2} \int d^4 x d^4 y d\tau \bar \eta (x, \tau) K^{-1} (x, y ) \eta (y, \tau) \right).    
 \end{equation}
After writing down the partition function, we can change $\bar \eta ,  \eta  $ to $\bar \psi, \psi$, and thus obtain 
\begin{eqnarray}
 Z &=& \int D \bar \psi D \psi  \det\left[\frac{\delta \bar \eta}{\delta \bar\psi}\right]^{-1}  
 \det\left[\frac{\delta  \eta}{\delta \psi}\right]^{-1} \nonumber \\ && \times 
  \exp \left( - \frac{1}{2} \int d^4 x d^4 y d\tau \bar \eta (x, \tau) K^{-1} (x, y ) \eta (y, \tau) \right). 
\end{eqnarray}
Now using Langevin equations, we obtain 
\begin{eqnarray}
 Z &=&  \int D \bar \psi D \psi  \det
 \left[K^{-1}\frac{\partial }{\partial \tau} - \frac{\delta^2 S[\bar\psi, \psi]}{ \delta \bar \psi \delta \psi } 
\right]^{-1}  
 \det \left[K^{-1}\frac{\partial }{\partial \tau} +  \frac{\delta^2 S[\bar\psi, \psi]}{ \delta  \psi \delta \bar \psi } 
\right]^{-1}\nonumber \\ && \times 
 \exp \left(- \frac{1}{2} \int d^4 x d^4 y d\tau \left[\frac{\partial \bar \psi}{\partial \tau} K^{-1} -
 \frac{\delta  S[\bar\psi, \psi]}{   \delta  \psi }  \right]\right. \nonumber \\ && \left. \times 
\left[\frac{\partial \psi }{\partial \tau}  +  K  \frac{\delta  S[\bar\psi, \psi]}{  \delta \bar \psi }  \right]
\right), 
\end{eqnarray}
where the determinant can be defined as the regularized product of eigenvalues. 
Thus, we can obtain an expression for 
the determinant  in terms of  ghost fields $ (c_1,\bar c_1 ,  c_2, \bar c_2 )$,  because we can write the following expression  \cite{11st}, 
\begin{eqnarray}
\bar c_1   \det \left[ K^{-1} \frac{\partial}{\partial \tau} - \frac{\delta^2 S}{\delta \psi \delta \bar \psi } \right]  c_1 = \lambda \bar c_1 c_1,  \nonumber \\ 
\bar c_2 \det \left[ K^{-1} \frac{\partial}{\partial \tau} - \frac{\delta^2 S}{\delta \psi \delta \bar \psi } \right]  c_2 = \lambda \bar c_2 c_2. 
\end{eqnarray}
 With this partition function, we can write 
\begin{eqnarray}
 \mathcal{L}_{eff} &=& \frac{1}{2}   \frac{\partial \bar \psi}{\partial \tau} K^{-1}\frac{\partial  \psi}{\partial \tau} -
 \frac{1}{2}  \frac{\delta  S[\bar\psi, \psi]}{   \delta  \psi } K  \frac{\delta S[\bar\psi, \psi]}{   \delta  \bar \psi }
\nonumber \\  && +  \bar c_1 
 \left[K^{-1}\frac{\partial }{\partial \tau} - \frac{\delta^2 S[\bar\psi, \psi]}{ \delta \bar \psi \delta \psi } 
\right]c_1  \nonumber \\  &&   + 
 \bar c_2  \left[K^{-1}\frac{\partial }{\partial \tau} +  \frac{\delta^2 S[\bar\psi, \psi]}{ \delta  \psi \delta \bar \psi } 
\right] c_2. 
\end{eqnarray}
We can introduce auxiliary fields $\bar F$ and $F$ and write the partition function for this theory as \cite{12st}-\cite{10st}
\begin{equation}
 Z = \int D\bar \psi D \psi Dc_1 D\bar c_1 Dc_2 D\bar c_2  D F D \bar F \exp\left( - \int d^4 x e  \mathcal{L}_{eff} \right), 
\end{equation}
where now 
\begin{eqnarray}
 \mathcal{L}_{eff} &=&  2 \bar F K^{-1} F + i 
 \left[\frac{\partial \bar \psi }{\partial \tau} K^{-1} - \frac{\delta  S[\bar\psi, \psi]}{ \delta \psi } 
\right] F \nonumber \\ &&
+  i 
\bar F  \left[\frac{\partial   \psi }{\partial \tau}   + K \frac{\delta  S[\bar\psi, \psi]}{ \delta \bar \psi } 
\right] +  \bar c_1 
 \left[K^{-1}\frac{\partial }{\partial \tau} - \frac{\delta^2 S[\bar\psi, \psi]}{ \delta \bar \psi \delta \psi } 
\right]c_1  \nonumber \\  &&   + 
 \bar c_2  \left[K^{-1}\frac{\partial }{\partial \tau} +  \frac{\delta^2 S[\bar\psi, \psi]}{ \delta  \psi \delta \bar \psi } 
\right] c_2. 
\end{eqnarray}

We can express this action using the superfield formalism. In order to do that, we define 
 superfields $\Omega (x, \tau, \theta, \bar \theta) $ and $\Omega (x, \tau, \theta, \bar \theta) $, as  
\begin{eqnarray}
 \Omega(x, \tau, \theta, \bar \theta)  &=& \psi (x, \tau) + \bar \theta c_1(x, \tau)  + \bar c_2(x, \tau)  \theta + i \theta \bar \theta F(x, \tau) ,
 \nonumber \\
  \bar \Omega(x, \tau, \theta, \bar \theta)  &=& \bar \psi(x, \tau)  + \bar \theta c_2(x, \tau)  + \bar c_1(x, \tau)  \theta + i \bar \theta
  \theta \bar F(x, \tau) .
\end{eqnarray}
We also define  superderivatives and   supercharges as 
\begin{eqnarray}
 D =  \frac{\partial }{\partial \bar \theta } - \theta \frac{\partial}{\partial \tau}, && Q =  \frac{\partial }{\partial \bar \theta }, 
 \nonumber \\
 \bar Q =  \frac{\partial}{\partial \theta} + \bar  \theta \frac{\partial}{\partial \tau}, &&\bar D = \frac{\partial}{\partial \theta}, 
\end{eqnarray}
 where   
\begin{eqnarray}
 \{ D, \bar D \} =  - \frac{\partial }{\partial  \tau}, &&
  \{ Q, \bar Q \} =  \frac{\partial }{\partial  \tau}. 
\end{eqnarray}
 The action  $ S = S_1 + S_2 + S_3$ coincides with the component action, if we make the following changes $ \bar F  K^{-1}\to \bar F, 
 \bar  \psi K^{-1} \to \bar \psi, \bar c_1 K^{-1} \to \bar c_1, \bar c_2 K^{-1} \to \bar c_2 $,  and 
 $F \to F, \psi \to \psi, c_1 \to c_1, c_2 \to c_2$ \cite{ferm}, where 
\begin{eqnarray}
 S_1 &=&  \int d^4 x e d\bar \theta d \theta \bar D \bar \Omega  D \Omega
 \\ \nonumber &=& \int d^4 xe  \left[ \bar c_1\frac{\partial c_1 }{\partial \tau}  + \bar F F + i \bar F \frac{\partial \psi}{\partial \tau}\right],
\\ \nonumber 
   S_2 &=&  \int d^4 x e d\bar \theta d \theta \bar D  \Omega  D \bar \Omega 
 \\ \nonumber &=& \int d^4 x e \left[ \bar c_2\frac{\partial c_2 }{\partial \tau}  +  F \bar F + i \bar F \frac{\partial \bar \psi}{\partial \tau}\right],
 \\ \nonumber 
 S_3 &=&  \int d^4 x e d\bar \theta d \theta S [ \bar \Omega, \Omega ]
 \\ \nonumber &=& \int d^4 x e \left[ \bar F \frac{\delta S_3}{\delta \bar \psi } +  \frac{\delta S_3}{\delta   \psi } F
 +   \bar c_1  \frac{\delta^2  S_3}{\delta \psi \delta \bar \psi } c_1  + \bar c_2 \frac{\delta^2  S_3}{\delta \psi  \delta \bar \psi  } c_2  \right]. 
\end{eqnarray}
Thus,  stochastic quantization of the deformed Dirac equation can be done using superspace formalism. 
 It may be noted that we have   performed the  stochastic quantization of the deformed Dirac equation using the semi-classical 
 approximation. Thus, during this analysis, we have taken the gravitational field as a fixed background field, and 
 performed the stochastic quantization of the deformed Dirac equation on this fixed background  field.

\section{Conclusion}

In this paper, we first clarified what it means to have a minimum measurable time in the theory. 
This was  done by first defining  the observable time  
as the time of occurrence of an certain event   at  which   
  some specified value gets assigned to a dynamical quantity  \cite{time}-\cite{time1}.
Then viewing minimum measurable time as the    existence of a  minimum measurable 
value to the certainty with which the  occurrence of that  event   can be measured by 
  the appearance of this specified value for that dynamical quantity. 
Then we  have combined the deformation  of the   Heisenberg algebra   caused by the  
  doubly special relativity and with its deformation caused by 
the existence of  minimum  measurable 
length and time scales.  We also have completed this deformed Heisenberg algebra by including a covariant 
 deformation of its temporal part. It may be noted that one of the most interesting consequence of the spatial part of this 
deformed Heisenberg algebra was that it 
 it leads to the discretization of space \cite{n4}. It will be interesting to repeat this analysis for the full Heisenberg algebra. 
 It is expected that such 
 an analysis will lead to the discretization of time.  
It was observed that this deformed algebra gives rise to non-local fractional derivative terms. However, a formal meaning was 
given to these non-local terms in the framework of harmonic extension of functions.
 This deformed Dirac equation was also analysed on curved spacetime. It was demonstrated that 
   this deformation of the Dirac equation also deforms the geometry of spacetime. Thus, corrections to the  Einsteins equations were obtained 
    by analysing this deformation. 
Finally, the 
stochastic quantization of this deformed Dirac equation was performed on curved spacetime.
It was done by using a fictitious time coordinate and 
anticommuting fermionic Gaussian noise. Thus, we were able to analyse the 
   Langevin equation for the deformed Dirac action on curved spacetime.
   It was also observed that the stochastic quantization of this deformed Dirac equations 
   could be done using superspace
formalism. It may be noted that we did not analyse the stochastic quantization of gravitational field in this model. 
   It would be interesting to analyse the stochastic quantization of this modified theory of gravity 
   in the framework of perturbative quantum gravity.
 However, as not all degrees of freedom in this modified theory of gravity  are physical, 
 we will need to add gauge fixing   and ghost terms,  
 to the original action. This resultant effective action,  which will be given by a sum of the original action with the gauge fixing and ghost terms, 
 will be invariant under a new symmetry called the BRST symmetry. In fact, various   
 consequences of the BRST symmetry of perturbative quantum gravity have been analysed \cite{grav1}-\cite{grav2}. It would be interesting to 
 analyse these consequences of the BRST symmetry for this modified theory of gravity.  
 
Non-local fractional derivative terms also occur in fermionic theories with Lifshitz scaling, and  
stochastic quantization of such Lifshitz theories  has also been performed \cite{ferm}.
It would thus be interesting to combine the effects coming from the deformation of the Dirac equation with Lifshitz scaling.
We expect that  non-local  fractional derivative terms,  that will occur due to the combination of the 
deformation of the Heisenberg algebra with Lifshitz scaling, could still be   analysed using the
harmonic extension of functions. It is also expected that by analysing the deformed Dirac equation with Lifshitz scaling,
further modification to the geometry will occur. It would be interesting to analyse this modification to the Einstein equations. 
Corrections to various cosmological models can be calculated from these modified Einstein equations. Furthermore,  
  phenomenological predications about cosmology can also be made from such  corrected cosmological models. It may be noted that the deformation 
of the Wheeler-DeWitt equation caused by this algebra has already been analysed \cite{mir2}.
It may be noted that there are various  proposals for the problem of time in quantum gravity \cite{ab}-\cite{cd}, 
and it would be interesting to analyse the deformation of models based on such proposals. 
Thus, it would also be interesting to analyse 
the modification to quantum cosmology that occurs because of this deformed algebra.
It may also be noted that an alternative deformation of the Heisenberg algebra can lead to noncommutative spacetime \cite{non1}-\cite{non2}. 
The noncommutativity has been generalized to non-anticommutativity, and perturbative 
  quantum gravity on non-anticommutative spacetime has   been analysed  \cite{qg}-\cite{qg1}. 
It would be interesting to combine the deformation of the Heisenberg algebra studied in this paper with non-anticommutativity, 
and analyse the effect of this combined deformation on quantum gravity. 
Recently, it has also been observed that all black objects 
in gravity's rainbow  will have a remnant \cite{gr}, and this has phenomenological consequences for detection of higher dimensional 
black holes at the LHC \cite{gr1}. As both gravity's rainbow   and the modification 
of gravity studied in this paper are motivated by doubly special relativity, it would be interesting to analyse black objects in 
the modified theory of gravity discussed in this paper. It is possible that a similar result may hold for all black objects in this
modified theory of gravity.

\end{document}